\documentclass[english,a4paper,twocolumn,amsmath,amssymb]{revtex4-1}
\usepackage{graphicx,color,soul}

\newcommand{\dd}{\mathrm{d}}

\newcommand{\pd}[2]{\frac{\partial #1}{\partial #2}}

\newcommand{\mean}[1]{\langle #1 \rangle}
\newcommand{\Int}[1]{\int\dd #1\;}
\newcommand{\IInt}[3]{\int_{#2}^{#3}\dd #1\;}

\renewcommand{\vec}[1]{\mathbf #1}

\newcommand{\gam}{\gamma}

\newcommand{\lam}{\lambda}

\newcommand{\sig}{\sigma}
\newcommand{\Sig}{\Sigma}

\newcommand{\kT}{k_\text{B}T}
\newcommand{\nois}{\boldsymbol\xi}
\newcommand{\x}{\vec r}
\newcommand{\z}{\omega}

\newcommand{\eq}{_\text{eq}}

\begin{document}

\title{Thermodynamic formalism for transport coefficients \\ with an application to the shear modulus and shear viscosity}

\author{Thomas Palmer}
\author{Thomas Speck}
\affiliation{Institut f\"ur Physik, Johannes Gutenberg-Universit\"at Mainz,
  Staudingerweg 7-9, 55128 Mainz, Germany}

\begin{abstract}
  We discuss Onsager's thermodynamic formalism for transport coefficients and apply it to the calculation of the shear modulus and shear viscosity of a monodisperse system of repulsive particles. We focus on the concept of extensive ``distance'' and intensive ``field'' conjugated via a Fenchel-Legendre transform involving a thermodynamic(-like) potential, which allows to switch ensembles. Employing Brownian dynamics, we calculate both the shear modulus and the shear viscosity from strain fluctuations and show that they agree with direct calculations from strained and non-equilibrium simulations, respectively. We find a dependence of the fluctuations on the coupling strength to the strain reservoir, which can be traced back to the discrete-time integration. These results demonstrate the viability of exploiting fluctuations of extensive quantities for the numerical calculation of transport coefficients.
\end{abstract}

\maketitle


\section{Introduction}

Transport coefficients determine the strength of currents in response to an external field (\emph{i.e.}, ``conductivities''), or vice versa (``resistivities''), in the linear response regime. A famous example is Ohm's law, which relates the current density to the electric field through the electrical conductivity. The determination of transport coefficients is an important aspect of computational materials science, intriguing examples include thermal transport in nanostructures~\cite{berb00,cahi03}, quantum liquids~\cite{marc16}, and two-dimensional materials~\cite{zera16}; and charge transport in disordered media~\cite{ruhe11,kord16}.

There are two fundamental possibilities for the calculation of transport coefficients. First, one can perform explicit non-equilibrium simulations in which a system is driven by an external field or non-conservative force. Measuring the average current as a function of driving strength then allows to extract the transport coefficient from the initial linear slope. Two drawbacks of this approach are: (i)~the dependence on the finite size of the simulated system, which often requires some sort of extrapolation in order to determine bulk properties~\cite{sche02,visc07} and (ii)~smaller fields lead to larger errors. In principle one can control either the field (``Th\'evenin'' ensemble) or the current (``Norton'' ensemble), but typically one of these two can be implemented more easily. For example, for Brownian dynamics and linear shear flow it is straightforward to add the velocity profile to the equations of motion and thus to simulate at constant strain rate (\emph{viz.} constant current). Simulating at constant stress might offer new physical insights~\cite{brow86,wang15,vezi15}, but requires some sort of feedback or ``constrained simulations''~\cite{morriss,evan86,stro16}. While equilibrium properties are defined by the Boltzmann factor and thus energies, the situation is different for dynamics breaking detailed balance since the characteristics of the ensuing non-equilibrium steady state do depend on these dynamics. This makes it desirable to not tamper with the dynamics.

The second approach is to exploit the fact that thermal fluctuations encode the response of a system to a small change of external conditions. This is well known from statistical mechanics, \emph{e.g.}, the specific heat, as a material property, determines the amount of heat flowing in response to a change of temperature. The extension to transport coefficients goes back to seminal work by Onsager~\cite{onsa31,onsa31a}, which is mostly known for his reciprocal relations. Arguably even more important is that this symmetry of transport coefficients is due to the existence of a potential, which allows to extent concepts from thermodynamics to the linear response regime (and even to non-equilibrium steady states~\cite{spec16b}). Through this potential one identifies pairs of conjugated intensive field (also called generalized force or affinity) $f$ and extensive ``distance'' $X$.

Despite Onsager's work, the prevalent view on transport coefficients is a dynamic one. The starting point is the explicit dynamics of the system, which is perturbed through switching on an external force. The linear deviation from equilibrium is calculated, which leads to Green-Kubo relations involving the integration over a two-point correlation function of fluxes~\cite{kubo,morriss}. In some cases these fluxes can be rewritten as a total derivative, which allows the Green-Kubo formula to be recast in terms of an Einstein-Helfand relation~\cite{helf60}. Transport coefficients can then be determined either from moments evaluated at explicit boundaries~\cite{petr05} or in bulk implementing periodic boundaries~\cite{visc03,visc07}.

Here we are interested in Onsager's thermodynamic formalism for systems that can be described through a coupling to an (effective) reservoir. The system is driven by ``draining'' the reservoir with a non-zero average rate of change for the extensive quantity. Specifically, we consider a simple model system for colloidal (nearly) hard spheres moving in a solvent with overdamped Brownian dynamics. This means that velocities remain equilibrated with the solvent temperature at all times even if the system is driven through shear flow. We illustrate the analogy with thermodynamics by first calculating the shear modulus for the solid at high densities, before going to lower densities at which the system fluidizes. The stress in a flowing suspensions is governed by the viscosity, which for a hard sphere fluid has been calculated first by Alder \emph{et al.} using the corresponding Helfand moment~\cite{alde70}. In contrast, we calculate the viscosity from strain fluctuations stemming from coupling the suspension to a strain reservoir characterized by a constant stress.


\section{Theory}
\label{sec:theory}

\subsection{Equilibrium}

It is instructive to recall how ensembles arise in statistical mechanics~\cite{chandler}. An isolated system corresponds to the microcanonical ensemble, for which the extensive quantities (energy $E$, volume $V$, particle number $N$, \dots) are conserved and thus fixed with thermodynamic potential $S(\{X_i\})$ (the microcanonical entropy). The derivatives
\begin{equation}
  \pd{S}{X_i} = f_i
\end{equation}
define the intensive generalized forces. A second ensemble is constructed through dividing the total system with fixed $X'$ into the system of interest holding a small amount $X$ and a \emph{reservoir} with remainder $X'-X$. Hence, $X$ is not conserved anymore but exchanged with the reservoir, which is an ideal construct with the property that the corresponding force $f$ remains constant.

Now consider an isolated system with two conserved quantities, the energy $E$ and some $X$ with entropy $S(E,X)$. Coupling to a heat reservoir, the Boltzmann factor for this canonical ensemble reads $\psi\eq(\z;\beta,X)=e^{-\beta E(\z)+\beta\mathcal F(\beta,X)}$ with energy $E(\z)$ depending on microstate $\z$, $f_E=-\beta$ the inverse temperature $\beta\equiv(\kT)^{-1}$ of the heat reservoir, and free energy $\mathcal F(\beta,X)$ arising from the normalization. Coupling to another reservoir (see Fig.~\ref{fig:res}), we obtain the Boltzmann factor
\begin{equation}
  \psi\eq(\z;\beta,f) = e^{-\beta E(\z)+fX(\z)+\beta\mathcal G(\beta,f)}
\end{equation}
with average
\begin{equation}
  \mean{X} = -\pd{(\beta\mathcal G)}{f}.
\end{equation}
Response relations are then given by
\begin{equation}
  \label{eq:susc:f}
  \pd{f}{X} = \pd{^2(\beta\mathcal F)}{X^2}
\end{equation}
and
\begin{equation}
  \label{eq:susc:X}
  \pd{\mean{X}}{f} = -\pd{^2(\beta\mathcal G)}{f^2}
  = \mean{X^2} - \mean{X}^2.
\end{equation}
In the thermodynamic limit ($N,V\to\infty$ with $N/V$ held fixed) the two thermodynamic potentials $\mathcal F$ and $\mathcal G$ are related through a Fenchel-Legendre transform~\cite{touc09}.

\subsection{Linear response regime}

\begin{figure}[b!]
  \centering
  \includegraphics{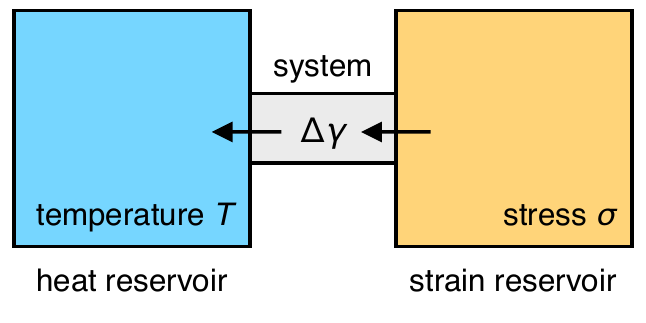}
  \caption{A system coupled to two reservoirs: a heat reservoir at temperature $T$ and a strain reservoir at stress $\sig$. If the system cannot sustain a finite stress (\emph{e.g.}, because it is a fluid) it will constantly remove strain $\gam$ from the reservoir, which we demand to remain at constant $\sig$. The strain reservoir thus performs work that is dissipated into the heat reservoir, with the system attaining a non-equilibrium steady state.}
  \label{fig:res}
\end{figure}

Now suppose that the coupled system and reservoirs do not reach equilibrium anymore but the system is ``draining'' one reservoir. This is what happens to the fluid coupled to a reservoir with non-zero stress, which starts to flow accumulating strain. The strain reservoir is thus spending work on the system, which is dissipated into the heat reservoir, see Fig.~\ref{fig:res}. The entropy production reads
\begin{equation}
  \label{eq:epr}
  \mathcal S = f\Delta X_t,
\end{equation}
whereby $\Delta X_t\equiv X_t-X_0$ is the change of the extensive quantity (a generalized distance) over time $t$. Note that $\Delta X_t\sim t$ is required to be time-extensive, \emph{i.e.}, the average increases linearly with time. We assume that a steady state is reached in which time-translational invariance holds.

An elegant way to derive the fluctuation-dissipation theorem is the fluctuation theorem, which quite generally states that $\mean{e^{-\mathcal S}}=1$~\cite{gall95,lebo99,seif12}. Expanding the exponential and truncating after the second order leads to $\mean{\mathcal S}=\frac{1}{2}\mean{\mathcal S^2}$~\cite{gall96,andr07b}. Inserting Eq.~(\ref{eq:epr}) we obtain
\begin{equation}
  \label{eq:fdt}
  \mean{\Delta X_t} = \frac{1}{2}\mean{(\Delta X_t)^2}\eq f + \dots
\end{equation}
where the right hand side is the lowest order (\emph{i.e.}, linear) contribution of the generalized force $f$, with the correlations thus to be evaluated in equilibrium. Here and in the following we assume that for $f=0$ the system reaches thermal equilibrium.

Now consider the probability distribution $P(\Delta X;t)$ of the change $\Delta X$ during time $t$. While the concrete form of this distribution depends on the dynamics, we do not have to specify the dynamics. We do, however, require the steady state distribution for a small force $f$ to be
\begin{equation}
  \label{eq:P:f}
  P(\Delta X;t) \approx \frac{e^{\frac{1}{2}f\Delta X}P\eq(\Delta X;t)}{Z(f;t)}
\end{equation}
with normalization
\begin{equation}
  \label{eq:Z:f}
  Z(f;t) \equiv \Int{X} e^{\frac{1}{2}f\Delta X} P\eq(\Delta X;t).
\end{equation}
Here, $P\eq(\Delta X;t)$ is the distribution in thermal equilibrium, which obeys the condition $P\eq(-\Delta X;t)=P\eq(\Delta X;t)$ since there is no transport in equilibrium with $\mean{\Delta X_t}\eq=0$. As a consequence, the distribution Eq.~(\ref{eq:P:f}) obeys the fluctuation theorem $P(\Delta X;t)/P(-\Delta X;t)=e^{\mathcal S}$.

The function $Z(f;t)$ [Eq.~(\ref{eq:Z:f})] has the form of a generating function, from which we can obtain moments by differentiation with respect to $f/2$. Drawing from the theory of large deviation, we consider the asymptotic limit of large times $t$ (as denoted by the symbol '$\asymp$') when both functions
\begin{equation}
  \label{eq:asymp}
  P\eq(\Delta X;t) \asymp e^{-t\Phi(J)/2} ~~\text{and}~~
  Z(f;t) \asymp e^{t\Phi^\ast(f)/2}
\end{equation}
decay as simple exponentials with current $J\equiv\Delta X/t$. In this limit they are related by the Fenchel-Legendre transform~\cite{touc09}
\begin{equation}
  \label{eq:legendre}
  \Phi^\ast(f) = \sup_J [fJ-\Phi(J)].
\end{equation}
On the right hand side we recognize Onsager's principle of least dissipation of energy~\cite{onsa31,onsa31a}, where $fJ=\mathcal S/t$ is the entropy production rate and $\Phi(J)$ is the dissipation function. Instead of just being the supremum (as in Onsager's original exposition), Eq.~(\ref{eq:legendre}) determines the large deviation function $\Phi^\ast(f)$. Combining Eqs.~(\ref{eq:Z:f}) and~(\ref{eq:asymp}), the first two derivatives become
\begin{equation}
  \mean{\Delta X_t} = t\pd{\Phi^\ast}{f}, \qquad
  \frac{1}{2}\mean{(\Delta X_t)^2}\eq\asymp t\left.\pd{^2\Phi^\ast}{f^2}\right|_{f=0}.
\end{equation}
Relating these expressions to the fluctuation-dissipation theorem Eq.~(\ref{eq:fdt}), we see that to lowest order the large deviation function can be written
\begin{equation}
  \Phi^\ast(f) = \frac{1}{2}\chi f^2
\end{equation}
with conductivity
\begin{equation}
  \label{eq:chi}
  \chi \equiv \lim_{t\to\infty}\frac{1}{2}\pd{}{t}\mean{(\Delta X_t)^2}\eq
  = \left.\pd{\mean{J}}{f}\right|_{f=0}.
\end{equation}
Note that this result is different from a typical Helfand moment~\cite{helf60} since it involves the total change $\Delta X$ of an additional degree of freedom (the extensive quantity exchanged with the reservoir). It extents Eq.~(\ref{eq:susc:X}), which also involves the extensive quantity. Both relations exploit the fluctuations in a finite system, highlighting that transport coefficients are indeed the analog of thermodynamic susceptibilities.


\section{Shear modulus}

\subsection{Constant strain}

To illustrate these ideas we now turn to a specific pair of conjugate observables: strain $\gam$ ($X\to V\gam$) and stress $\sig$ ($f\to\beta\sig$). We start by considering a system in thermal equilibrium that can sustain a finite stress, which implies that the system is in the solid state. Throughout we consider the canonical ensemble with fixed number of particles $N$ and fixed volume $V$ with density $\rho=N/V$. A microstate $\z\equiv\{\x_i\}$ is specified by the knowledge of all particle positions $\x_i=(x_i,y_i,z_i)^T$. The system has potential energy $U(\z)$ and can exchange energy with a heat reservoir at temperature $T$. In addition, the shape of the bounding box might be deformed. We only consider shear but more general deformations can be implemented easily~\cite{parr81}. In the following, we omit the kinetic part of the energy since it is independent of the deformation. Such an affine deformation is characterized by the strain $\gam$ and preserves the total volume $V$. We orient the coordinate system so that the new particle positions $\x_i\to\x^\gam_i$ after the deformation read
\begin{equation}
  \label{eq:aff}
  \x^\gam_i = \x_i + \gam y_i\vec e_x
\end{equation}
with potential energy $U(\z,\gam)=U(\z^\gam)$.

Keeping the strain $\gam$ fixed, the partition function reads
\begin{equation}
  Z(\gam) = \frac{1}{\lam_T^{3N}} \sum_\z e^{-\beta U(\z,\gam)} = e^{-\beta\mathcal F(\gam)}
\end{equation}
with thermal de Broglie wavelength $\lam_T$. The derivative of the free energy $\mathcal F(\gam)$ leads to
\begin{equation}
  \label{eq:F}
  \frac{1}{V}\pd{\mathcal F}{\gam}
  = \frac{1}{V}\sum_\z\pd{U}{\gam} \frac{1}{Z}e^{-\beta U(\z,\gam)}
  = \mean{\hat\sig(\z^\gam)}_\gam
\end{equation}
with \emph{microscopic} shear stress
\begin{equation}
  \label{eq:str}
  \hat\sig(\z) \equiv \frac{1}{V} \sum_{i=1}^N y_i\pd{U}{x_i}
  = \frac{1}{V}\sum_{i<j} \frac{x_{ij}y_{ij}}{r_{ij}} u'(r_{ij})
\end{equation}
assuming pairwise interactions, $U(\z)=\sum_{i<j}u(|\x_i-\x_j|)$. Here, the sum runs over all pairs of particles with separation $\x_{ij}\equiv\x_i-\x_j$ and $r_{ij}=|\x_{ij}|$. The brackets $\mean{\cdot}$ in Eq.~(\ref{eq:F}) denote the thermal average, whereby the subscript emphasizes that the strain is kept constant.

The shear modulus
\begin{equation}
  \label{eq:G}
  G_\gam \equiv \pd{\mean{\hat\sig}}{\gam}
  = \frac{1}{V}\pd{^2\mathcal F}{\gam^2}
\end{equation}
is a thermodynamic susceptibility (or response coefficient), cf. Eq.~(\ref{eq:susc:f}). Performing the differentiation leads to explicit expressions for the shear modulus that have be used in computer simulations~\cite{hess97,fara00}. The result 
\begin{equation}
  \label{eq:G:fluct}
  G_\gam = G^\text{BG} - \Sigma_\gam
\end{equation}
can be split into two contributions, the Born-Green term $G^\text{BG}$ and the fluctuations
\begin{equation}
  \label{eq:var:gam}
  \Sigma_\gam \equiv \beta V\left(\mean{\hat\sig^2} - \mean{\hat\sig}^2\right)_\gam
\end{equation}
of the microscopic stress. One should note that $G^\text{BG}$ and the stress fluctuations are typically evaluated for the unstrained solid ($\gam=0$).

\subsection{Strain reservoir}

Eq.~(\ref{eq:G:fluct}) is often sufficient to calculate the shear modulus in computer simulations. Consequently, the fact that one can use
\begin{equation}
  \pd{\mathcal F}{(V\gam)} = \sig
\end{equation}
to change the ensemble $\gam\to\sig$ so that the stress is the controlled variable has been explored to a much lesser degree. We now consider the situation that the system is coupled to a reservoir (see Fig.~\ref{fig:res}) with which it can exchange strain (analogous to the heat reservoir with which it exchanges energy). Note that in the $\sig$-ensemble the microscopic stress $\hat\sig$ measured in a finite system is \emph{not} constant but fluctuates around its average $\mean{\hat\sig}_\sig=\sig$.

Following Sec.~\ref{sec:theory}, the partition function and free energy read
\begin{equation}
  Z(\sig) = \frac{1}{\lam_T^{3N}} \sum_{\z,\gam} e^{-\beta[U(\z,\gam)-V\gam\sig]}
  = e^{-\beta\mathcal G(\sig)},
\end{equation}
where $\gam$ is now an additional degree of freedom. It is straightforward to check that the average strain
\begin{equation}
  \mean{\gam}_\sig = -\frac{1}{V}\pd{\mathcal G}{\sig}
\end{equation}
follows as expected with susceptibility [cf. Eq.~(\ref{eq:susc:X})]
\begin{equation}
  \label{eq:G:sig}
  \frac{1}{G_\sig} = \pd{\mean{\gam}}{\sig} 
  = \beta V\left(\mean{\gam^2}-\mean{\gam}^2\right)_\sig.
\end{equation}
The final relation with the strain fluctuations follows immediately from the fact that $Z(\sig)$ is not only the partition function but also the generating function of the strain statistics. It states that the fluctuations of the strain encode its response to an external infinitesimal change of the stress. One should note that the reverse does not hold, \emph{i.e.}, $Z(\gam)$ is not the generating function for fluctuations of $\hat\sig$, which is a function of the microstate. Indeed, while superficially similar, Eq.~(\ref{eq:G:fluct}) follows for $\gam=0$, involves an additional contribution $G^\text{BG}$, and the fluctuations have to be subtracted. Nevertheless, there is a famous result by Lebowitz, Percus, and Verlet~\cite{lebo67} that allows the express correlation functions in a different ensemble \emph{in the thermodynamic limit}. Applying this relation to the stress fluctuations leads to
\begin{equation}
  \label{eq:lpv}
  \Sigma_\sig \equiv \beta V\left(\mean{\hat\sig^2} - \mean{\hat\sig}^2\right)_\sig
  = \Sigma_\gam + \pd{\mean{\hat\sig}}{\gam} = G^\text{BG}
\end{equation}
after inserting Eq.~(\ref{eq:G:fluct}), which predicts that the stress fluctuations in the $\sig$-ensemble are given by the Born-Green term (which itself is ensemble-independent).

\subsection{Brownian dynamics}
\label{sec:bd}

While so far everything has been independent of the actual dynamics, we now restrict our attention to Brownian dynamics
\begin{equation}
  \label{eq:bd}
  \dot\x_{it} = -\mu_0\nabla_iU(\z_t,\gam_t) + \sqrt{2\mu_0\kT}\nois_{it}
\end{equation}
with normal, independent noise $\nois_i$ and bare mobility $\mu_0$. Eq.~(\ref{eq:bd}) describes the evolution of the \emph{reference} microstate $\z_t$ but employing the potential $U(\z,\gam)$ depending on the instantaneous strain. For the time evolution of the strain we use
\begin{equation}
  \dot\gam_t = \frac{\beta V}{\tau}[\sig-\hat\sig(\z^\gam_t)] + \xi^\gam_t
\end{equation}
with noise correlations
\begin{equation}
  \mean{\xi^\gam_t\xi^\gam_{t'}} = \frac{2}{\tau}\delta(t-t'),
\end{equation}
where the time constant $\tau$ is, in principle, a free parameter determining the coupling to the strain reservoir. It quantifies the relaxation of the microscopic stress, with small $\tau$ leading to fast relaxation and concurrently large strain fluctuations (strong coupling), and large $\tau$ leading to slow relaxation and small strain fluctuations (weak coupling).

From these stochastic equations of motion we derive the time evolution of the joint probability $\psi(\z,\gam,t)$,
\begin{multline}
  \label{eq:2}
  \pd{\psi}{t} = \mu_0\sum_{i=1}^N\nabla_i\cdot[(\nabla_iU)+\kT\nabla_i]\psi + \\ \frac{1}{\tau}\pd{}{\gam}\left[-\beta V(\sig-\hat\sig)+\pd{}{\gam}\right]\psi.
\end{multline}
With 
\begin{equation}
  \pd{\psi\eq}{\gam} = \beta V[-\hat\sig(\z^\gam)+\sig]\psi\eq(\z,\gam)
\end{equation}
the stationary solution ($\partial_t\psi=0$) becomes the Boltzmann factor
\begin{equation}
  \label{eq:eq}
  \psi\eq(\z,\gam) = \frac{1}{Z(\sig)} e^{-\beta[U(\z,\gam)-V\gam\sig]},
\end{equation}
which is independent of $\tau$. The corresponding equations of motion for deterministic Hamiltonian dynamics are given in Appendix~\ref{sec:det}.

\subsection{Simulations}

\begin{figure}[b!]
  \includegraphics{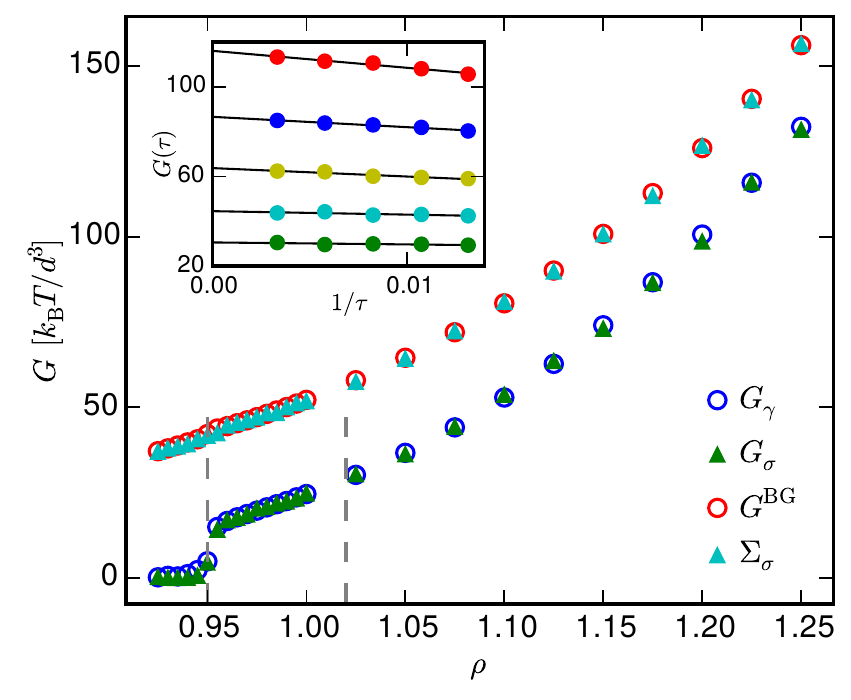}
  \caption{Shear modulus of the WCA solid as a function of density $\rho$ for $T=1$. The dashed lines indicate the freezing and melting densities bounding the fluid-solid coexistence. Shown are: $G_\gam$ (circles) from simulations at fixed $\gam$ straining the solid and $G_\sig$ (triangles, extrapolated to $\tau\to\infty$) calculated from strain fluctuations [Eq.~(\ref{eq:G:sig})], both of which agree excellently. The inset shows the dependence of $G_\sig(\tau)$ on the inverse coupling time $1/\tau$ with linear fits for five exemplary densities. We have also calculated the Born-Green term $G^\text{BG}$ (circles) and the fluctuations $\Sig_\sig$ (triangles) of the microscopic stress in the $\sig$-ensemble (again $\tau\to\infty$). Both agree as predicted by Eq.~(\ref{eq:lpv}). Errors are smaller than symbol sizes.}
  \label{fig:modulus}
\end{figure}

To test the predictions made for the various relations between the two ensembles, we have performed computer simulations. As a concrete model system, we study a one-component system of particles interacting via the Weeks-Chandler-Andersen (WCA) potential~\cite{week72}
\begin{equation}
  u(r) = \begin{cases}
    4\epsilon\left[(d/r)^{12} - (d/r)^{6}\right] + \epsilon & (r/d<2^{1/6}) \\
    0 & (r/d\geqslant2^{1/6}),
  \end{cases}
\end{equation}
which is the purely repulsive part of the Lennard-Jones potential shifted by its minimum value. We simulate at least $N=1000$ particles with periodic boundaries. The phase diagram is basically that of hard spheres with a fluid phase at low density, fcc solid at high density, and a coexistence region~\cite{ahme09}. In the following we employ dimensionless quantities with energies measured in units of $\epsilon$, lengths measured in units of $d$, and time measured in units of $d^2/(\mu_0\epsilon)$. The equations of motion are integrated with a finite time step $\delta t=10^{-4}$ ($\delta t=10^{-5}$ for the stress fluctuations $\Sig_\sig$). All simulations were initialized on an fcc lattice and relaxed for $\approx0.5$ Brownian times before data recording was begun.

The simulation results are summarized in Fig.~\ref{fig:modulus}. First, from simulations with fixed strain we obtain $G_\gam$ as the slope from linear fits of $\mean{\hat\sig}$ as a function of (small) strain $\gam$. These values match the data of Hess \emph{et al.}~\cite{hess98}. We then performed simulations in the $\sig$-ensemble (as described in the previous section) with $\sig=0$, recording the strain $\gam$ and calculating its variance, from which we obtain $G_\sig$ through Eq.~(\ref{eq:G:sig}). While we have shown that $\tau$ drops out for the continuous equations of motion, we do observe a dependence of $G_\sig$ on $\tau$ for our simulation results. This can be traced back to the discretization of the equations of motion with a finite time step, see Appendix~\ref{sec:db} for details. Agreement with $G_\gam$ should be recovered in the limit $\delta t/\tau\to0$ (\emph{i.e.}, weak coupling), which is indeed the case as demonstrated in Fig.~\ref{fig:modulus}. To this end, we have performed several simulations varying $\tau$ and extrapolating to $1/\tau\to0$. We have also determined the variance of the microscopic stress $\hat\sig$, which agrees with the Born-Green contribution to the shear modulus as expected.

\begin{figure}[t]
  \includegraphics{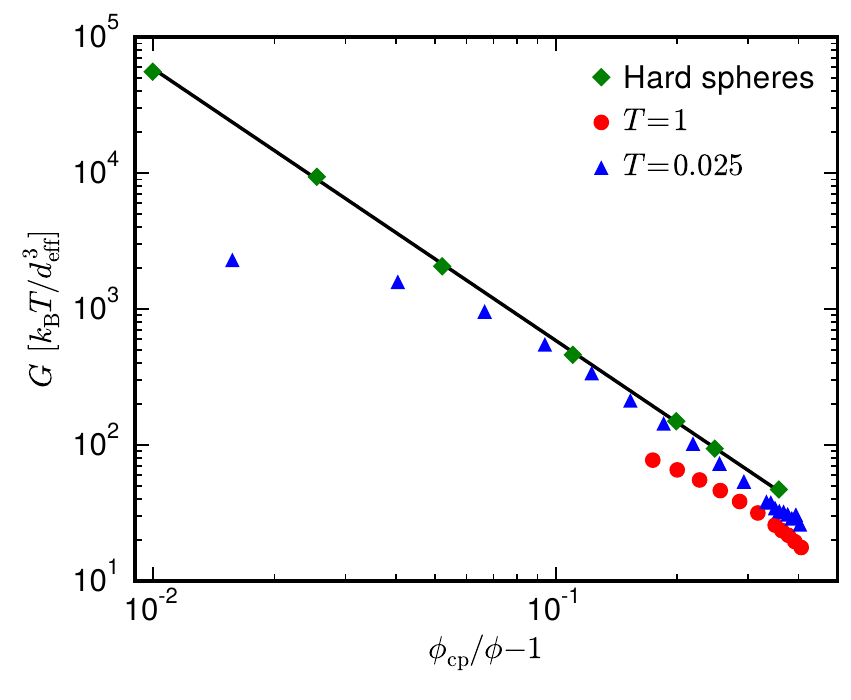}
  \caption{Shear modulus $G_\gam$ of the WCA solid for temperatures $T=1$ (discs) and $T=0.025$ (triangles) compared to true hard spheres [data from Farago and Kantor~\cite{fara00}, solid line corresponds to Eq.~(\ref{eq:HS})]. The higher temperature results considerably deviate from this conjectured form of the shear modulus. At $T=0.025$, the WCA particles follow the line (with constant offset) until $\phi_\text{cp}/\phi-1\approx 0.07$, which is equivalent to $\phi\simeq0.69$ ($\rho\simeq1.32$).}
  \label{fig:hard}
\end{figure}

For completeness, we compare our WCA solid with true hard spheres. To this end, we determine a temperature dependent effective diameter $d_\text{eff}$~\cite{bark67} with volume fraction $\phi=\rho\pi d_\text{eff}^3/6$. Simple arguments predict the form
\begin{align}
  G^\text{HS} = \frac{A}{\phi_\text{cp}/\phi -1} \frac{\kT}{d^3}
  \label{eq:HS}
\end{align}
for the shear modulus of hard spheres with diameter $d$, which has been reproduced using Monte Carlo simulations with constant $A\simeq 5.86$~\cite{fara00}. Here, $\phi_\text{cp}\simeq0.74$ is the close-packing fraction of equal spheres. In Fig.~\ref{fig:hard} it is shown that at low temperature $T=0.025$ the WCA solid indeed follows this form for not too high densities (with a slightly lower value for $A$) but deviates strongly at high densities due to the higher compressibility of the softened interactions.


\section{Shear viscosity}

\subsection{Stress fluctuations}

\begin{figure}[b!]
  \includegraphics{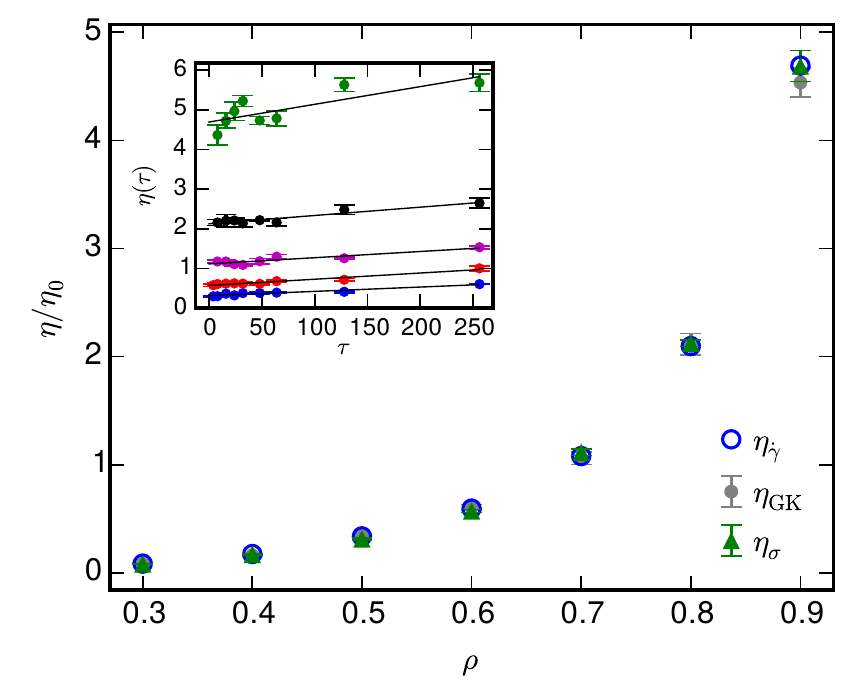}
  \caption{Shear viscosity $\eta$ (due to interactions) of the WCA fluid as a function of density $\rho$ for $T=1$ ($\eta_0$ is the solvent viscosity). Shown are results using three different methods: $\eta_{\dot\gam}$ (open circles) from direct non-equilibrium simulations at fixed strain rate $\dot\gam$, $\eta_\text{GK}$ (discs) from the Green-Kubo relation [Eq.~(\ref{eq:eta:gk})], and $\eta_\sig$ (triangles) from the strain fluctuations via Eq.~(\ref{eq:eta:fluct}) in the limit $\tau\to0$ of infinitely fast stress relaxation. Inset: Dependence of $\eta_\sig$ on the coupling time $\tau$ with linear fits used for the extrapolation.}
  \label{fig:visc}
\end{figure}

We now consider the identical WCA system but at densities below the freezing density, $\rho<0.95$. Since then it is a fluid, the system's response to a non-vanishing stress is to flow. In virtually all simulations of this situation the strain rate $\dot\gam$ is controlled and the average stress $\mean{\hat\sig}$ in the fluid is measured, from which the shear viscosity follows as
\begin{equation}
  \label{eq:eta}
  \eta_{\dot\gam} = \left.\pd{\mean{\hat\sig}}{\dot\gam}\right|_{\dot\gam=0}.
\end{equation}
This expression corresponds to Eq.~(\ref{eq:susc:f}) for the solid. Note that it takes into account only the contribution due to the conservative forces leading to the stress Eq.~(\ref{eq:str}). We have performed non-equilibrium simulations with an explicit linear shear flow
\begin{equation}
  \dot\x_{it} = \dot\gam y_{it}\vec e_x - \mu_0\nabla_iU(\z_t) + 
  \sqrt{2\mu_0\kT}\nois_{it}
\end{equation}
using Lees-Edwards boundary conditions~\cite{allen}. We have calculated the stress and from the initial slope one can extract $\eta_{\dot\gam}$, which is shown in Fig.~\ref{fig:visc} as a function of density.

From the explicit stochastic dynamics, it is straightforward to derive the Green-Kubo relation
\begin{equation}
  \label{eq:eta:gk}
  \eta_\text{GK} = \beta V\IInt{t}{0}{\infty} 
  \mean{\hat\sig(\z_t)\hat\sig(\z_0)}\eq,
\end{equation}
which expresses the viscosity as an integral over the time correlations of the microscopic stress $\hat\sig(\z)$. The simulation results from equilibrium runs in the $\gam$-ensemble with $\gam=0$ are shown in Fig.~\ref{fig:visc} and agree very well with $\eta_{\dot\gam}$.

\subsection{Strain fluctuations}

\begin{figure}[t]
  \centering
  \includegraphics{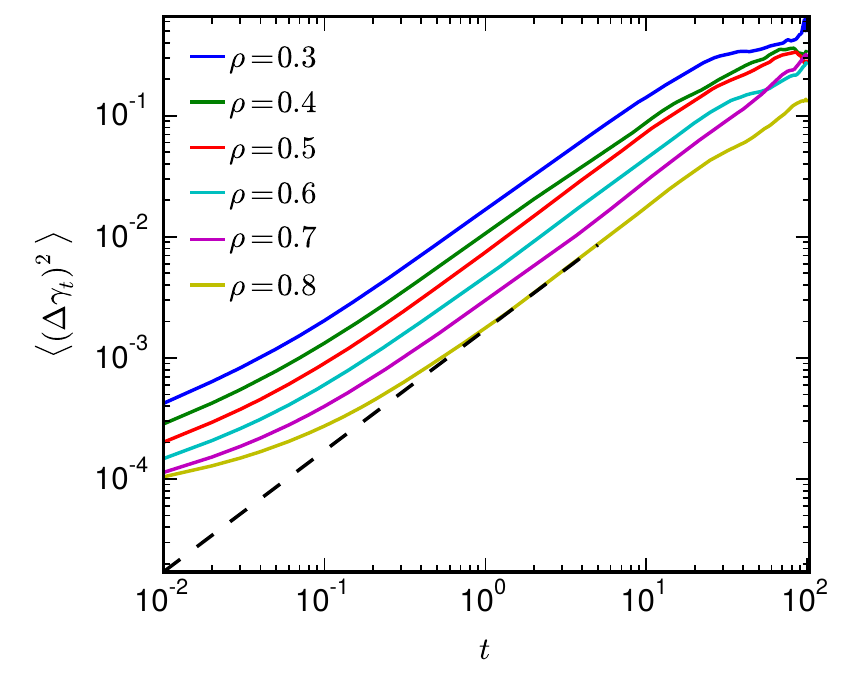}
  \caption{Strain fluctuations $\mean{(\Delta\gam_t)^2}$ as a function of time $t$ for fixed $\tau=8$. After a transient the fluctuations increase proportional to $t$. At long times reaching the total length of trajectories, the statistics breaks down. We fit the linear regime for $t<5$, see example fit for $\rho=0.8$ (dashed line).}
  \label{fig:fluct}
\end{figure}

Changing the ensemble, we perform further equilibrium simulations but now in the $\sig$-ensemble with the equations of motion given in Sec.~\ref{sec:bd}. While the strain $\gam$ now fluctuates, at $\sig=0$ its mean still vanishes, $\mean{\gam}=0$. In contrast, the variance $\mean{(\Delta\gam_t)^2}$ of the strain change $\Delta\gam_t\equiv\gam_t-\gam_0$ increases, see Fig.~\ref{fig:fluct}. After a short transient, this increase becomes proportional to $t$. Eq.~(\ref{eq:chi}) now becomes
\begin{equation}
  \label{eq:eta:fluct}
  \frac{1}{\eta_\sig} = \pd{\mean{\dot\gam}}{\sig} = \lim_{t\to\infty}\frac{\beta 
    V}{2}\pd{}{t}\mean{(\Delta\gam_t)^2}\eq
\end{equation}
and predicts that the linear slope is related to the inverse viscosity. This relation is confirmed in Fig.~\ref{fig:visc}, showing that the viscosities $\eta_\sig$ thus determined agree with the other two approaches. While the coupling time $\tau$ should not influence the stationary distribution, in the simulations we again observe a dependence on $\tau$, but now the correct value is obtained in the limit $\tau\to0$ of strong coupling. Finally, we set $\sig>0$ to a non-zero but small value and indeed observe that the suspension starts to flow with non-zero $\mean{\dot\gam}>0$.

\section{Conclusions}

We have numerically tested an alternative approach to determining transport coefficients, which is based on the fluctuations of \emph{extensive} quantities exchanged with reservoirs. Here, we have studied the shear modulus and shear viscosity of a model system and found excellent agreement with other numerical methods. This thermodynamic formalism allows to calculate transport coefficients from fluctuations in a rigorous way without perturbing the dynamics, and without introducing constrained dynamics~\cite{stro16} or unphysical forces/moves (swapping momenta, displacing particles, \dots) as is done in non-equilibrium molecular dynamics~\cite{mull99,jung16}. No assumption has to be made for the dynamics, which can be either stochastic or deterministic. Moreover, the thermodynamic approach avoids problems associated with defining the local fluxes for Green-Kubo relations, \emph{e.g.}, due to ambiguities in splitting of pairwise energies~\cite{sche02,marc16}, in defining the stress tensor~\cite{murd07}, and in the combination of stochastic dynamics with conserved quantities~\cite{espa95,erns06,espa09}. In the future it will be interesting to explore the connection of the response theory used here to the response in ensembles of trajectories with respect to integrated order parameters~\cite{jack16}. Another aspect is the transformation of shear-stress autocorrelation functions between ensembles, which has been discussed recently by Wittmer \emph{et al.}~\cite{witt15}.

For overdamped dynamics using a simple Euler-Maruyama integration scheme, the strain fluctuations do depend on the coupling strength to the reservoir and thus require extrapolation. This problem can be alleviated by employing a higher-order integration scheme. Interestingly, we observe that the two opposing limits corresponding to weak and strong coupling have to be used for the shear modulus and shear viscosity, respectively. Our results should be seen as a proof of principle, for the simple system studied here no performance gain is achieved, quite in constrast, the accurate determination of $\mean{(\Delta\gam)^2}$ is computationally expensive. Still, the generalization to heat and charge fluctuations is straightforward and the ideas used here might be beneficial, \emph{e.g.}, for the study of thermoelectric materials and nanoscale capacitors~\cite{limm13}.


\acknowledgments

We thank R.L. Jack and F. Schmid for helpful comments. The DFG is acknowledged for financial support through the collaborative research center TRR 146 (project A7). Computations were carried out on the Mogon Cluster at ZDV Mainz.


\appendix

\section{Deterministic dynamics}
\label{sec:det}

For the coupling to a strain reservoir we do not have to employ Brownian dynamics. For completeness, here we provide the deterministic equations of motion, on which molecular dynamics simulations are typically based (see also Ref.~\citenum{parr81} for the Lagrangian equations for more general deformations). The Hamiltonian of the system then reads
\begin{equation}
  H = \frac{1}{2}\sum_{i=1}^N\frac{\vec p_i^2}{2m_i} + U(\z,\gam)
  + \frac{\pi^2}{2M} - V\gam\sig,
\end{equation}
where $m_i$ is the mass and $\vec p_i$ is the momentum of particle $i$, and $\pi$ is the auxiliary ``momentum'' of the strain with coupling constant $M$. In equilibrium, the probability distribution is given by the Boltzmann factor $\psi\eq\sim e^{-\beta H}$, which reduces to Eq.~(\ref{eq:eq}) after integrating out all quadratic momenta. The equations of motion are
\begin{equation}
  \label{eq:det}
  \dot\x_i = \pd{H}{\vec p_i} = \frac{\vec p_i}{m_i}, \qquad
  \dot{\vec p}_i = -\pd{H}{\x_i} = -\nabla_iU
\end{equation}
and
\begin{equation}
  \label{eq:det:gam}
  \dot\gam = \pd{H}{\pi} = \frac{\pi}{M}, \qquad \dot\pi = -\pd{H}{\gam} = -V\hat\sig + V\sig
\end{equation}
with microscopic stress $\hat\sig$ defined in Eq.~(\ref{eq:str}). It is straightforward to check that the Poisson brackets $\{H,\psi\eq\}=0$ vanish, which guarantees that $\psi\eq$ is a constant of motion and that we sample the correct fluctuations of the strain $\gam$. This also extends to the case when thermostatting the dynamics Eqs.~(\ref{eq:det}). Note that Eqs.~(\ref{eq:det:gam}) are different from employing a simple ``feedback'' equation for the strain~\cite{evan86},
\begin{equation}
  \dot\gam = \frac{\beta V}{\tau}(\sig-\hat\sig),
\end{equation}
which leads to non-Liouvillian evolution and an unphysical stationary distribution of $\gam$ depending on $\tau$.

\section{Detailed balance}
\label{sec:db}

To understand the origin of the dependence of the shear modulus in the $\sig$-ensemble on the coupling strength $\tau$, consider the transition probability
\begin{align}
  p(\z\to\z') \propto \exp\left\{-\sum_i\frac{1}{4D_i\delta t}[x'_i-x_i-\delta t
  g_i(\z)]^2\right\}
\end{align}
between two configurations during the time step $\delta t$. The sum runs over all $3N+1$ degrees of freedom $\{x_i\}$ including the strain $\gam$, with equations of motion $\dot x_i=g_i+\sqrt{2D_i}\xi_i$ and independent Gaussian noises $\xi_i$.

The ratio of forward and backward transitions reads
\begin{multline}
  \frac{p(\z\to\z')}{p(\z'\to\z)} = \exp\bigg\{\sum_i\frac{1}{4D_i\delta
    t}\big(2\delta t(x'_i-x_i)[g_i(\z)+g_i(\z')] \\
  -(\delta t)^2[g_i(\z)^2-g_i(\z')^2]\big) \bigg\}.
\end{multline}
To first order, we thus obtain
\begin{align}
  \exp\left\{ -\frac{U(\z')-U(\z)}{\kT} + \delta
    t\sum_i\frac{1}{4D_i}[g_i(\z')^2-g_i(\z)^2] \right\},
\end{align}
which shows that in the simulations detailed balance is fulfilled for a ``shadow'' energy that reduces to $U(\z)$ in the limit $\delta t\to0$.  Inserting $g_\gam=(\beta V/\tau)(\sig-\hat\sig)$ and $D_\gam=1/\tau$, we see that for $\gam$ this extra contribution scales as $\delta t/\tau$.


%

\end{document}